\documentclass[3p,times,procedia]{elsarticle}
\usepackage{nupha_ecrc}
\usepackage{here}

\volume{00}

\firstpage{1}

\journalname{Nuclear Physics A}

\runauth{P.B. Gossiaux and R. Katz}

\jid{nupha}

\jnltitlelogo{Nuclear Physics A}

\usepackage{amssymb}

\usepackage[figuresright]{rotating}

\begin{document}

\begin{frontmatter}



\dochead{}

\title{Upsilon suppression in the Schr\"odinger-Langevin approach}


\author{P.B. Gossiaux$^1$ and R. Katz$^1$}

\address{$^1$SUBATECH, UMR 6457, Ecole des Mines de Nantes, Universit\'e de Nantes, IN2P3/CNRS.\\ 4 rue Alfred Kastler, 44307 Nantes cedex 3, France}

\begin{abstract}
We treat the question of bottomonia suppression in ultrarelativistic heavy ion collisions (URHIC) as a dynamical open quantum problem, tackled for the first time using the Schr\"odinger-Langevin equation. Coupling this equation to the EPOS2 event generator, predictions are made for the nuclear modification factor of $\Upsilon (1S)$ and $\Upsilon (2S)$.
\end{abstract}


\end{frontmatter}



\section{Motivation and model for $Q\bar{Q}$ dynamics}

Quarkonia production in URHIC -- see \cite{SaporeGravis:2015} for a recent review -- is one of the best probes of the color-deconfined QCD medium of high temperature -- the quark-gluon plasma (QGP) -- achieved in these collisions. The production of bottomonia is of particular interest as it is much less impacted by exogenous recombination than charmonia production and then less difficult to model. A dynamical calculation of the $b\bar{b}$ evolution satisfying the basic principles of quantum mechanics is a genuine open quantum system. As a possible method to tackle this problem, we use the so-called Schr\"odinger-Langevin equation (SLE) introduced by Kostin~\cite{KOstin:1972} for the dynamics of the internal degrees of freedom (relative coordinates of the pair)
\begin{equation}
i\hbar\frac{\partial \psi}{\partial t}= \Bigg[ \hat{H}_0 + \hbar A\Big(S({\bf x},t)-\int\psi^*S({\bf x},t)\,\psi \,\,d{\bf x }\Big) -
{\bf x}\cdot{\bf F}_R(t)\Bigg]\,\psi\,.
\label{SLeq}
\end{equation}
In this equation, the non linear term $\propto \hbar A$ is a dissipative friction term based on the wave function phase $S$ while the term $-{\bf x}\cdot{\bf F}_R(t)$ represents a stochastic dipolar force. The hamiltonian $\hat{H}_0$ encodes the dynamics in the absence of such interactions with the medium. The SLE admits some important properties: norm conservation -- despite a non linear term and contrarily to similar treatments relying on imaginary potentials --,  respect of the Heisenberg principle, 
violation of the superposition principle,\ldots The stochastic forces drive the system from a pure state to a mixed state. Observables are built using an ensemble average over these forces, which necessitates to evaluate 
 replications of the time evolution, a feature well suited to Monte Carlo simulations. The SLE naturally allows to deal with the transitions from the bound states to the continuum, at the core of quarkonia deconfinement.  
The asymptotic thermalization of a subsystem coupled to a heat bath is a particular issue of the SLE, whose mastering is essential to obtain reliable predictions. In \cite{Katz:2015qja}, we have investigated this problem for the 1D case, both for a harmonic and a linear potential and for white -- where $\langle F_R(t) F_R(t+\tau)\rangle_{\rm stat}= B \delta(\tau)$ -- and colored stochastic forces. Even though the state weights $W_n$ (defined as $W_n(t)=\langle | \langle n|\psi(t)|^2\rangle_{\rm stat}$) do not systematically converge towards some Boltzmann distribution $\exp(-E_n/T)$ for the "canonical" expression of $B$ -- $B=2 m A E_0 \left[\coth\left(\frac{E_0}{T}\right)-1\right]$ --, it is indeed the case for a subset of low lying eigenstates provided one adapts the autocorrelation $B$ as 
\begin{equation}
B=2 m A \tilde{T}(T)\,,
\label{eq:B_rescaling} 
\end{equation}
where the relation $\tilde{T}(T)$ depends on the particular potential.

For a first application of the SLE to the problem of bottomonia suppression in URHIC, we adopt a 1D modeling of the $b\bar{b}$ pairs, with even (resp.~odd) 1D-states mocking S (resp.~P) 3D-states. In vacuum, the potential in $\hat{H}_0$ is taken as $V_{\rm vac}(x)=K|x|$, truncated to $V_{\rm max}=1.2~{\rm GeV}$ (according to lQCD calculations \cite{Mocsy:2008}). $m_b$ is chosen as 4.575 GeV, while the string parameter $K = 1.375~{\rm GeV/fm}$ is chosen to obtain an energy difference between the first two even states $E_2-E_0 = E(\Upsilon')-E(\Upsilon) = 563~{\rm MeV}$. Even and odd bound states are found for 
$E=9.41, 9.97, 10.33$ GeV and $E=9.74, 10.18$ GeV which are in good agreement with corresponding experimental values for $\Upsilon(nS)$ and
$\chi_b(nP)$. One notices however that the binding energy of the $\Upsilon(3S)$ is just 20 MeV in this model, much smaller than the real 3D-case. Thus, little confidence is put into predictions concerning this precise state, a default that will be corrected in a future 
parametrization of the potential. Investigating the time evolution asymptotics with 
$V_{\rm vac}$ in $\hat{H}_0$ as well as with the friction term and the stochastic forces turned on, a detailed study shows that a correct thermalization is achieved by using $\tilde{T}(T)=0.83 T -0.08~{\rm GeV}$ in the expression
(\ref{eq:B_rescaling}) of $B$.
At finite temperature, however, the potential should be Debye-screened. This feature is implemented in our model by letting the maximal value $V_{\rm max}$ be $T$-dependent in $\hat{H}_0$ -- which then acquires the status of a mean field (MF) Hamiltonian -- and taken as the asymptotic ($r\rightarrow +\infty$) value of the finite-$T$ lQCD potential. As some ambiguity subsists on the most appropriate choice at finite-$T$, two prescriptions have been investigated for $V_{\rm max}$: the internal energy $U$~\cite{Kaczmarek:2005} (which shows some extra binding around $T_c$) as well as a potential suggested by Mocsy and Petreczky~\cite{Mocsy:2008} (our privileged choice in these proceedings) characterized by some weaker binding as compared to $U$. The corresponding 1D-potentials are referred to as $U$ and $V_W$. In the SLE approach, the friction/drag coefficient $A$ for the relative motion is considered to be identical to the one for single heavy flavor (a feature that can be intuited starting from the 2-body SLE and introducing the center of mass and the relative coordinates). It is taken from \cite{Gossiaux:1998} -- $A_b({\bf p}=0,T)
\approx 0.46 T + 0.32 T^2~{\rm (c/fm)}$ --  applying on the top a factor $K=1.5$ to match the open heavy flavor experimental data at LHC.

\section{Bottomonia suppression in a stationary medium}
\begin{figure}[H]
 \centering
\includegraphics[width=0.9\textwidth]{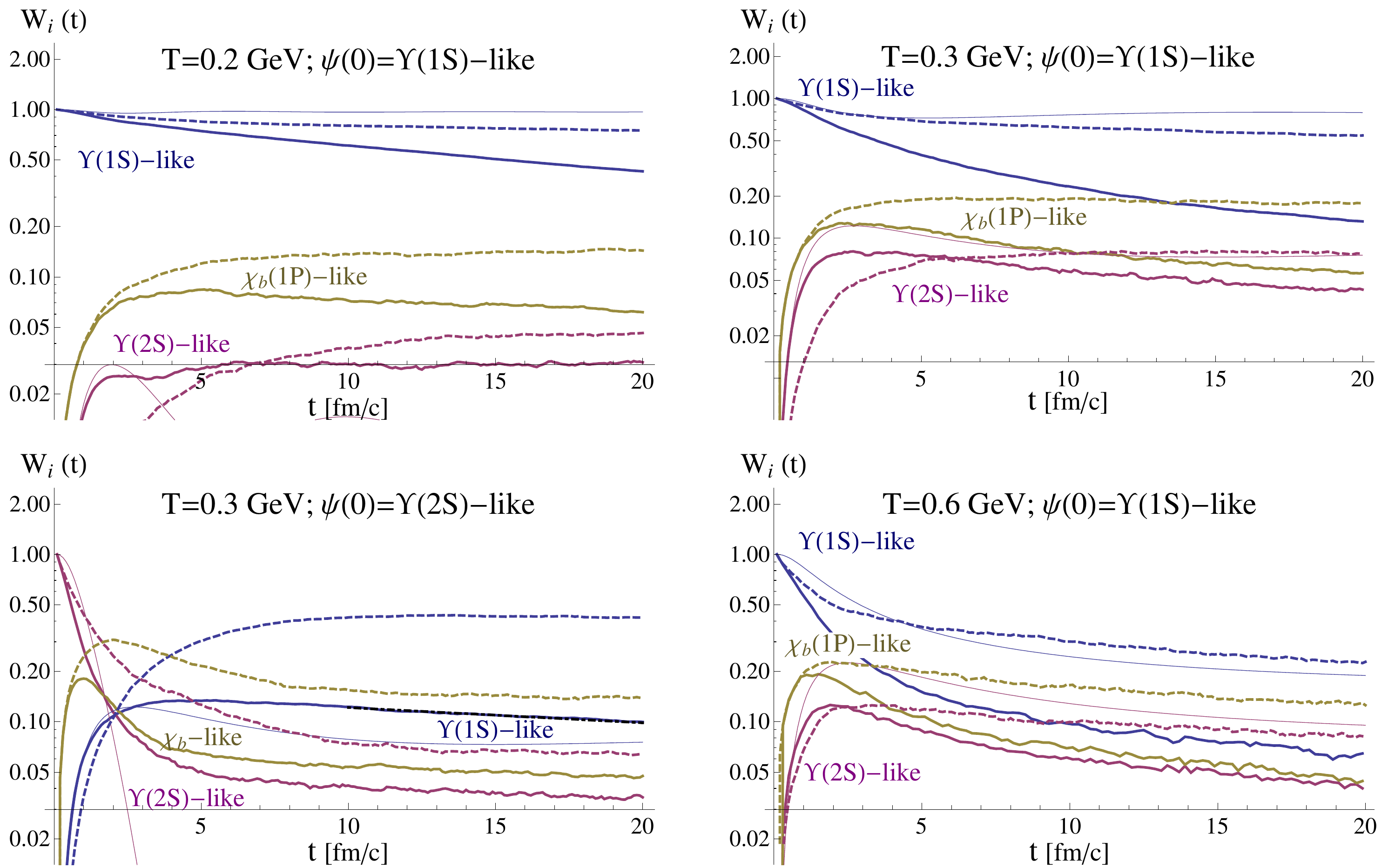} 
 \caption{(Color online) Evolution of the weights of the 3 lowest lying vaccum eigenstates (blue: 1S-like, brown: 1P-like, purple: 2S-like) in a stationary medium. Solid lines correspond to $V_W(T)$ without friction nor stochastic forces, while dashed and thick lines correspond respectively to $V_W(0)$ and $V_W(T)$ with friction and stochastic forces.}
 \label{fig:station}
\end{figure}

Contrarily to some common analysis, we do not pay much attention the precise temperatures at which the various bound states "melt"
under the influence of the MF in $\hat{H}_0$. We believe that such an analysis is not very relevant for the problem of bottomonia suppression in URHIC, not only because the medium is not stationary (see section \ref{sec:coll_URHIC}) but also because a) the evolution time of the $b\bar{b}$ is finite so that $b$ and $\bar{b}$ quarks which are found close to each other at the end of the evolution have a possibility to recombine into some bound state and b) if one waits
long enough, the stochastic forces may destroy bound states at any temperature anyhow. Instead, we investigate the capability of these forces as well as of the "screening" to modify the (vacuum) state content of some initial wave function $\psi(t=0)$ through dynamical evolution. 
We provide such an analysis in fig.~\ref{fig:station}, for typical QGP temperatures happening in URHIC and starting from a pure (vacuum) eigenstate considered at rest. For each temperature and each initial state, three models are investigated to understand the respective contributions of the screening and of the stochastic forces.
In all cases, the time evolution populates a large variety of $b\bar{b}$ eigenstates (at the exception of the evolution with the MF only which conserves the parity). When stochastic forces are turned on, one
observes, after some transient phase lasting $\propto A^{-1}$, an asymptotic regime where all considered weights decay uniformly. This is due to the global probability flow driven out of the potential well by these forces (while for the pure MF case, a finite probability is trapped in this well). During this asymptotic regime, the weight ratios tend towards values which solely depend on $T$ (as can be
seen by comparing both evolutions at $T=0.3~{\rm GeV}$, starting either 
from a 1S or a 2S state). We therefore interpret the transient phase as a re-equilibration of the $b\bar{b}$ d.o.f. leading to a mixed quantum state in thermal equilibrium with the surrounding QGP. From the comparison between the solid and the dashed curves on fig.~\ref{fig:station}, we realize that the stochastic forces are more efficient in destroying the $\Upsilon$-like state at small $T$ and the screening tend to me more efficient at high $T$. The largest suppression is obtained by combining both effects.

\section{Bottomonia suppression in URHIC}
\label{sec:coll_URHIC}
\begin{figure}[H]
 \centering
\includegraphics[width=0.45\textwidth]{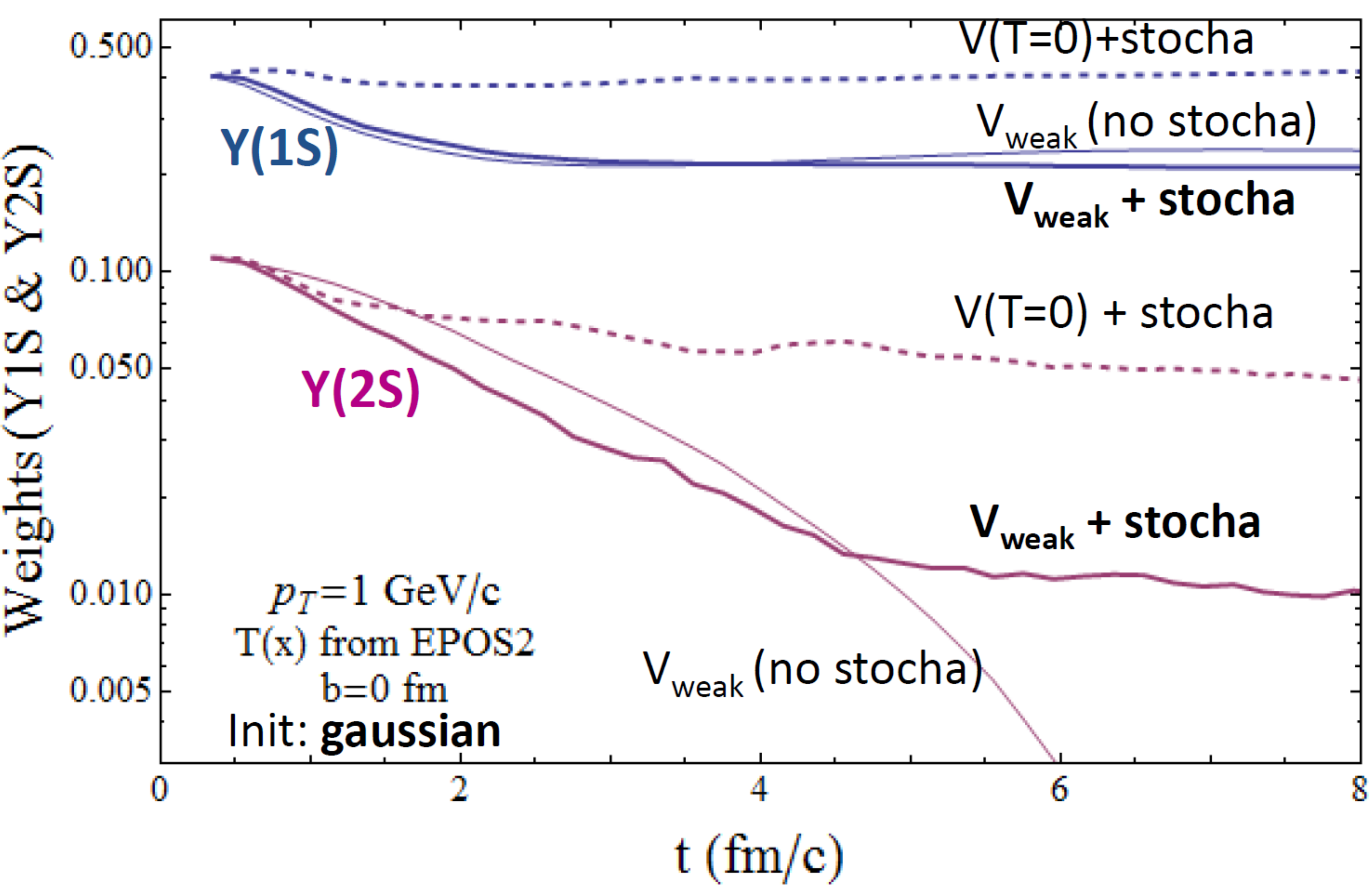} 
\hspace{5mm}
\includegraphics[width=0.43\textwidth]{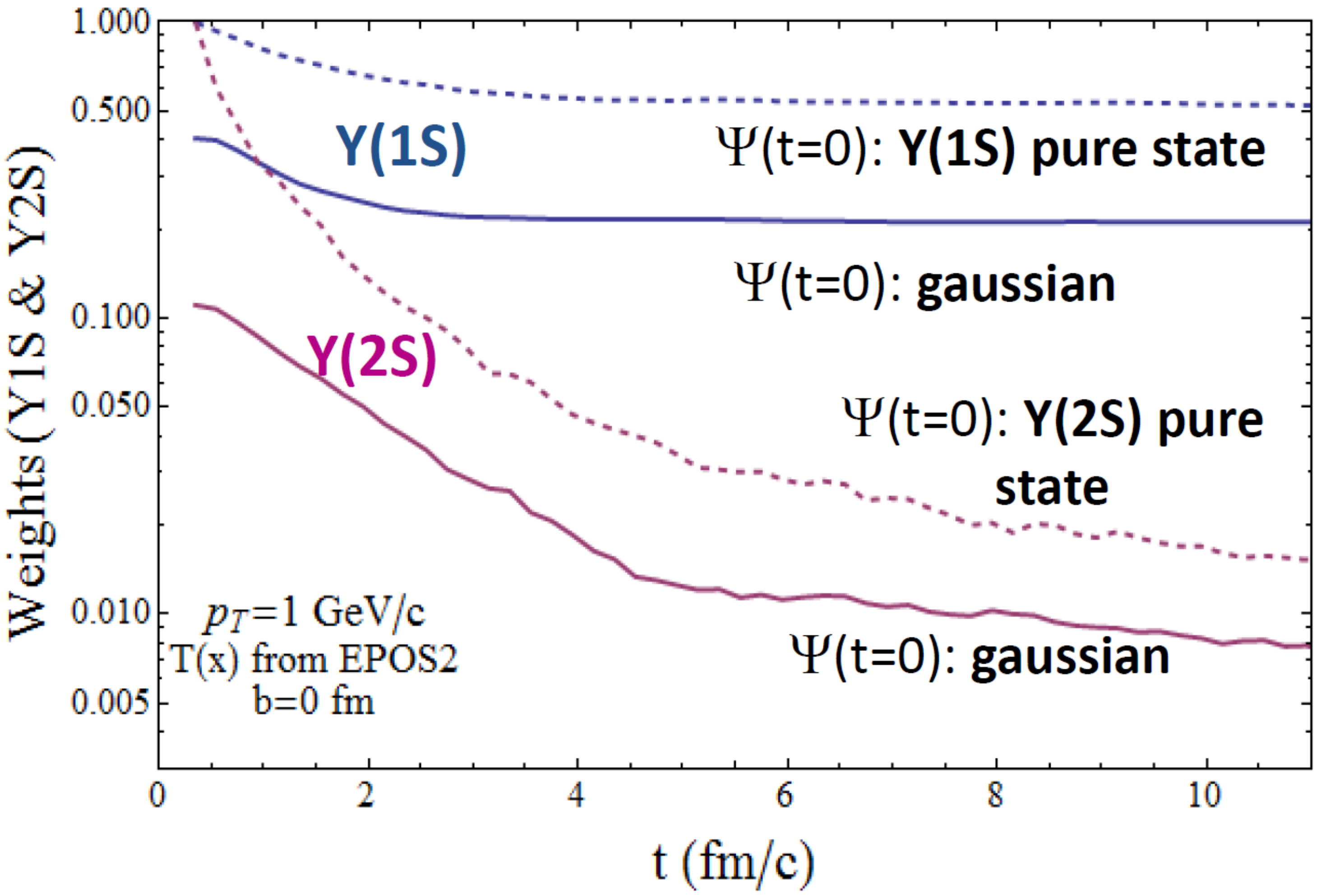} 
 \caption{(Color online) time evolution of the average $\Upsilon$ content of
 $b\bar{b}$ state at small $p_{\rm CM,T}$ for different variations of the SLE approach embedded in the EPOS2 profile (see text for details).}
 \label{fig:EPOS_evol}
\end{figure}

\begin{figure}[H]
 \centering
\includegraphics[width=0.45\textwidth]{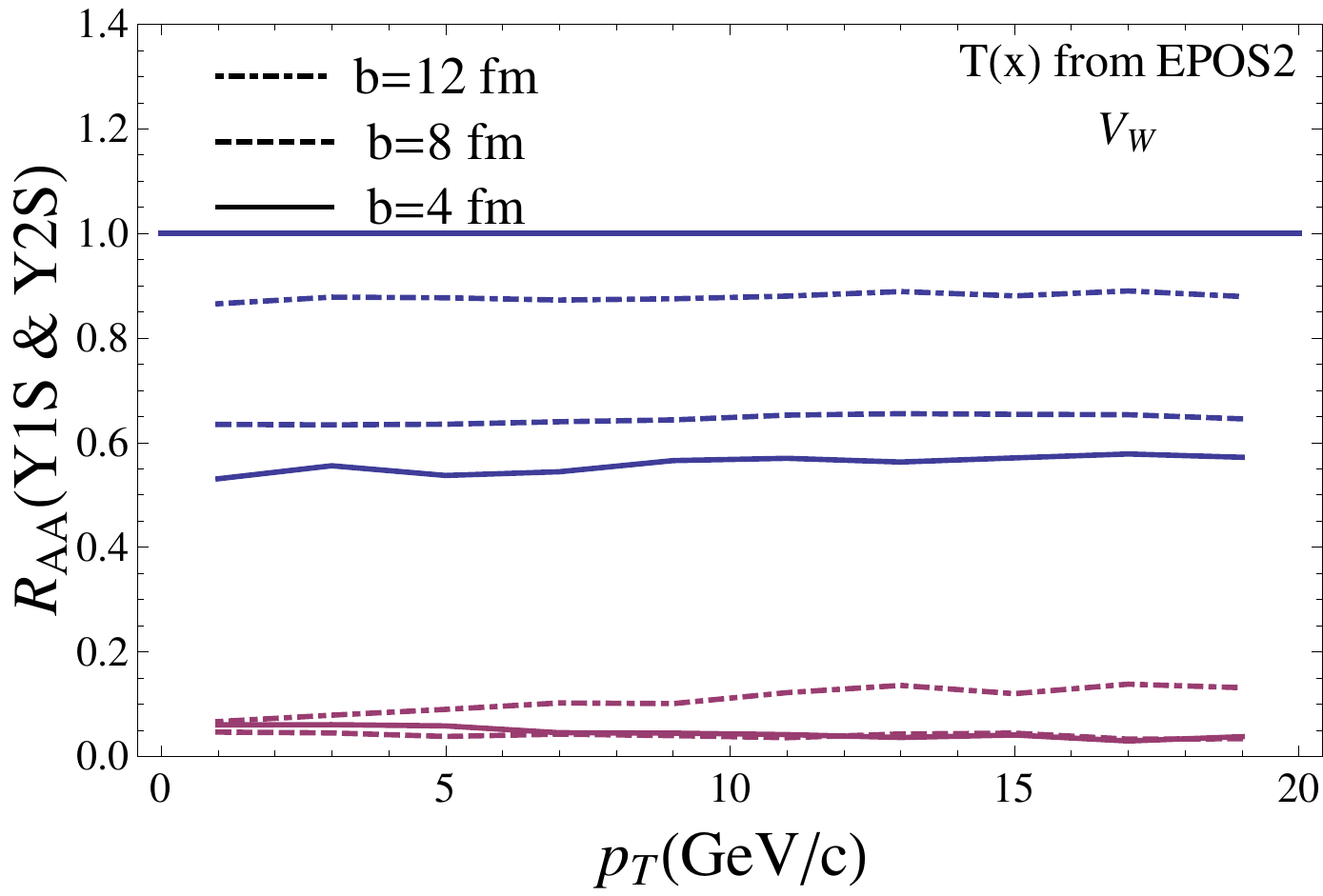} 
\hspace{5mm}
\includegraphics[width=0.43\textwidth]{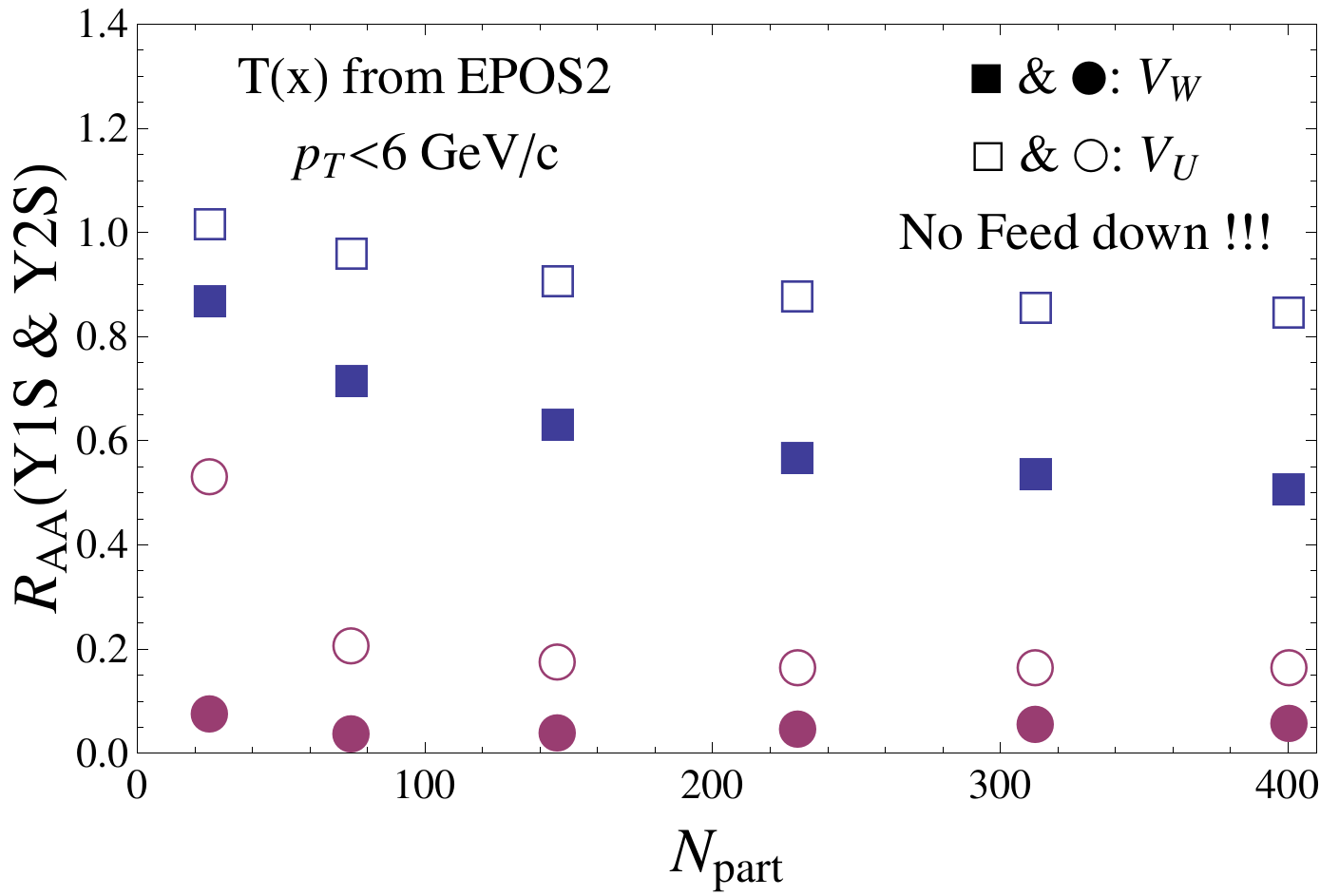} 
 \caption{(Color online) Nuclear modification factor of $\Upsilon$ and
 $\Upsilon'$ as a function of $p_T$ (left) and of $N_{\rm parr}$ (right).}
 \label{fig:RAA}
\end{figure}

We now consider dynamical $b\bar{b}$ in genuine URHIC, modeled by ideal fluid dynamics ensuing EPOS initial conditions~\cite{EPOS:2001,EPOS:2010} (hereafter named as EPOS2). For this purpose, we generate the initial positions of the $b\bar{b}$ pairs according to the Glauber model. Essential $b\bar{b}$ contributing to the production of final bound states are assumed to be global color-neutral states throughout the full evolution. The original center of mass momentum ${\bf p}_{\rm CM}$ is thus preserved and $b\bar{b}$ pairs follow straight trajectories along which the temperature entering the SLE ingredients is obtained from EPOS2 profiles, while the friction coefficient is evaluated from~\cite{Gossiaux:1998} at ${\bf p}_{b/\bar {b}}={\bf p}_{\rm CM}/2$. Typical time evolution are shown in fig.~\ref{fig:EPOS_evol}. In the left panel, we show the same information as in fig.~\ref{fig:station}, albeit with a more realistic initial $b\bar{b}$ state -- a Gaussian of width $0.045~{\rm fm}$, adjusted to the $\Upsilon'/\Upsilon$ ratio in pp. Contrarily to the stationary medium case, one observes a saturation of the weights at large time, corresponding to the disappearance of the screening and to the freezing of the thermal forces. Focusing on the $\Upsilon(1S)$, the evolution seems to be driven by the MF only, a feature that is obviously not generic (see fig.~\ref{fig:station}) and deserves
further investigation. For excited states, the  role of stochastic forces is crucial, as one can judge from the evolution of the $\Upsilon(2S)$-like content: Whereas these forces have the tendency to reduce it up to $t\approx 5~{\rm fm/c}$, they lead to a significant content afterwards (as compared to the pure MF evolution). This can be interpreted as a continuous repopulation of the $\Upsilon(2S)$-like content from the lowest lying states, an effect that is usually not correctly implemented in other approaches. On the right panel of fig.~\ref{fig:EPOS_evol}, we explore the role of the initial state on the weights  of $\Upsilon(nS)$ and on the associated "survival" $S_n$, defined as the ratio $W_n(t)/W_n(0)$. We notice, in particular, that $S_2$ is much {\em smaller} if one initiates the $b\bar{b}$ state as a pure 2S state. This is due, once again, to the repopulation of higher states from the 1S if one starts from a Gaussian wave packet. In fig.~\ref{fig:RAA}, we present these survivals $S_\Upsilon$ and $S_{\Upsilon'}$ at the last stage of the evolution -- which correspond, in  
our model, to the nuclear modification factor $R_{AA}$ measured in experiments -- both as a function of bottomonia transverse momentum $p_T$ and as a function of the number of participant in the $AA$ collision. Global trends are found to be in good agreement with experimental measurements~\cite{CMS:2015}, in particular the incomplete suppression of the $\Upsilon$ ground state as well as the remarkably flat $p_T$ dependence. 
  
\section{Conclusion}
To conclude, this first study shows the promising features of the Schr\"odinger-Langevin equation applied to the topic of bottomonia suppression in URHIC. Notice that no feed down has been included in the present work, a lack which will be corrected in an upcoming review paper containing more detailed analysis.

\section*{Acknowledgement}
We gratefully acknowledge the support from the TOGETHER project, R\'egion Pays de la Loire (France).





\bibliographystyle{elsarticle-num}



\end{document}